*Original Article*

# Outsourced Analysis of Encrypted Graphs in the Cloud with Privacy Protection

D. Selvaraj[1], S. M. Udhaya Sankar[2], D. Dhinakaran[3], T. P. Anish[4]

[1]Department of Electronics and Communication Engineering, Panimalar Engineering College, Chennai, India.
[2,3]Department of Information Technology, Velammal Institute of Technology, Chennai, India.
[4]Department of Computer Science and Engineering, R.M.K College of Engineering and Technology, Chennai, India.

[1]Corresponding Author : drdselva@gmail.com

*Abstract* - Huge diagrams have unique properties for organizations and research, such as client linkages in informal organizations and customer evaluation lattices in social channels. They necessitate a lot of financial assets to maintain because they are large and frequently continue to expand. Owners of large diagrams may need to use cloud resources due to the extensive arrangement of open cloud resources to increase capacity and computation flexibility. However, the cloud's accountability and protection of schematics have become a significant issue. In this study, we consider calculations for security savings for essential graph examination practices: schematic extraterrestrial examination for outsourcing graphs in the cloud server. We create the security-protecting variants of the two proposed Eigen decay computations. They are using two cryptographic algorithms: additional substance homomorphic encryption (ASHE) strategies and some degree homomorphic encryption (SDHE) methods. Inadequate networks also feature a distinctively confidential info adaptation convention to allow the trade-off between secrecy and data sparseness. Both dense and sparse structures are investigated. According to test results, calculations with sparse encoding can drastically reduce information. SDHE-based strategies have reduced computing time, while ASHE-based methods have reduced stockpiling expenses.

## 1. Introduction

The fast development of electronic gadgets and communications technology encourages the emergence of the cloud computing era, which has significant implications for and value for people from all walks of life. Data exporting services are accelerated by cloud computing, making them an essential and practical use [1]. The graph structure is common in many disciplines, including chemical structure, transportation, and sociological graphs. The cloud computing service often accepts extensive graph data, which is in charge of preserving, organizing, and analyzing such facts due to the cloud's tremendous processing capacity as well as the issue of cost savings [2]. However, because the CCP server is not entirely honest and reliable, it is important to consider and deal with the concerns about privacy related to the outsourced data. Before being outsourced to CCP, encryption of the exporting graph data is an efficient and often utilized technique [3]. However, encrypted outsourced graph data is difficult for data users to edit and use further. Implementing privacy-guarding optimum route discovery with assistance for query expansion on the secured graph in the cloud services context is thus critical work.

The privacy-preserving solutions of the majority of mining algorithms, particularly spectral information, can be built using two widely used general privacy-preserving techniques: homomorphic encryption and secure multi-party communication. However, their price makes them impractical [4]. Huge block cipher and intensive homomorphic multiplication are the consequences of the finest FHE scheme execution currently available. For recommender systems with scrambled networks, one cycle of factorization is a comparatively tiny $100 \times 100$ matrices using a confidentiality vector space algorithm in transmission.

Distributing graph data while protecting privacy is somewhat connected to our work. Its problem situation is quite different, in any case. To share graph data, publications must overcome privacy issues brought up by intrepid data miners [47]. With prior knowledge, however, the assaults can indeed be fully identified and comprehended. As a result, privacy for analyses has gained popularity recently. The publication of graph layouts is not our goal. Simple, sparse encoding, therefore, reveals excessive data because most graph matrices have unique structures. Our approach is the same as adding fictitious edges to achieve differential privacy [6]. The new entries are encoded 0s; therefore, they have no impact on the calculation of the matrix, so the reliability is unaffected by this edge insertion.

## 2. Related work

Unfortunately, gathering and processing graph data via the cloud raises privacy issues. Individuals are reluctant to provide these datasets since they are typically sensitive because they need to have faith in the ability of the data proprietors to keep the data source safe in the cloud server. On the other hand, as data are now crucial to doing business or conducting a scientific study, data owners also have a tremendous stake in maintaining their ownership of these valuable data. Furthermore, according to recent research and events, sensitive data stored in the cloud is vulnerable to data loss, spying, and malicious insiders. They are finding ways to accommodate consumers' and data proprietors' worries in cloud-based data extraction.

To assure security, several matrix processing methodologies have indeed been presented. These secured outsourced alternatives are tailored for large-scale linear regression solutions and applications involving multiplication and additive noise filtering. Their methods could be more effective because they reveal sensitive data, rely on numerous servers that are not collaborating, or need significant overhead. Use client-cloud cooperation and matrix disruption to solve systems of equations iteratively.

R. Bost [7] builds three main categorization protocols—decision trees, hyperplane decisions, and Naive Bayes —that satisfy this privacy restriction. They also make it possible for these methods to work with AdaBoost. They show that such libraries can also be utilized to design other predictors, such as multiplexing and feature extraction. These constructions are based on new libraries of essential components for reliably generating classifiers. They applied filters and libraries into practice and evaluated them. When used with actual clinical data, the efficient methods accomplish a diagnosis in a few milliseconds to a few seconds.

By fusing a customer's query information with permission data credentials and indices, D. Leilei [52] presents a Dynamic Multi-client SSE (DMSSE) method with support for boolean queries. The system restricts a client's search capability to appropriate terms and enables a data owner to authorize numerous clients to run boolean inquiries over an encrypted format. The advantages of our DMSSE scheme over current MSSE solutions include the following: 1) Lack of interaction. After receiving search authorization, clients are free to do their searches without the assistance of the data owner. 2) Active. The data holder can effectively change the search authorization of a customer without impacting other customers. Using the DMSSE method in a large encoded file is beneficial, as shown by empirical assessments performed on actual data.

Li et al. [9] presented a dynamic additive homomorphic encryption scheme and discussed a couple of crucial dilemmas using attribute-based encryption and the k-nearest neighbor algorithms. However, none of the searchable encryption alternatives can be employed to accomplish optimized route discovery with assistance for information retrieval over cryptographic graph data. F. Berger [48] to discover an implied representation of a molecule's ring system. They offer effective cyclic graph referential integrity techniques that could speed up lookups by acting as molecular descriptors. The precise construction of a molecular graph's well-defined collection of rings is yet another task. They provide a brand-new approach for calculating a graph's relevant cycle set.

Catalano, D [11] demonstrate a method for converting linearly homomorphic encryption into a system that can assess degree-2 calculations on encrypted message. The translation is remarkably easy to implement and only necessitates one very minor requirement on the baseline continuously homomorphic scheme: the communication field should be a public ring that allows evenly distributed sampling of its members. With practically all current number-theoretic linearly elliptic curve schemes, including Goldwasser-Micali, Paillier, or ElGamal, they can instantiate the transformations as a result. When addressing a subset of degree-2 harmonics in which the amount of modifications of degree-2 terms is constrained by a fixed, our resultant techniques ensure circuit confidentiality and are small. Z. Cui [12] concentrates on a fundamental issue with geo-tagged data: identifying the top k frequently occurring phrases in a particular region of the cloud's spatial data. They first create a Region Tree Index (RTI) for geo-tagged data. Then, Sorted Terms and Weights (SSTW) are suggested to be stored in RTI using the array collection architecture. The top k often occurring phrases in a specific area are computed using an effective k Terms Search method. Finally, thorough tests confirm the viability of the suggested scheme.

In a cloud computing context, Xianyi [13] provides a method to carry out privacy-protecting optimum route discovery with assistance for semantic search on the encrypted graph (PORF). Based on the concept of searchable encryption and the stemming process, we developed a method by creating a safe query index to execute optimum path discovery with assistance from the keyword web. For our system, they provide a rigorous security analysis. Furthermore, through experiments, they also evaluate the plan's effectiveness. GOOSE, a safe architecture for Graph Contracting and SPARQL Analysis, is presented by R. Ciucanu [14].

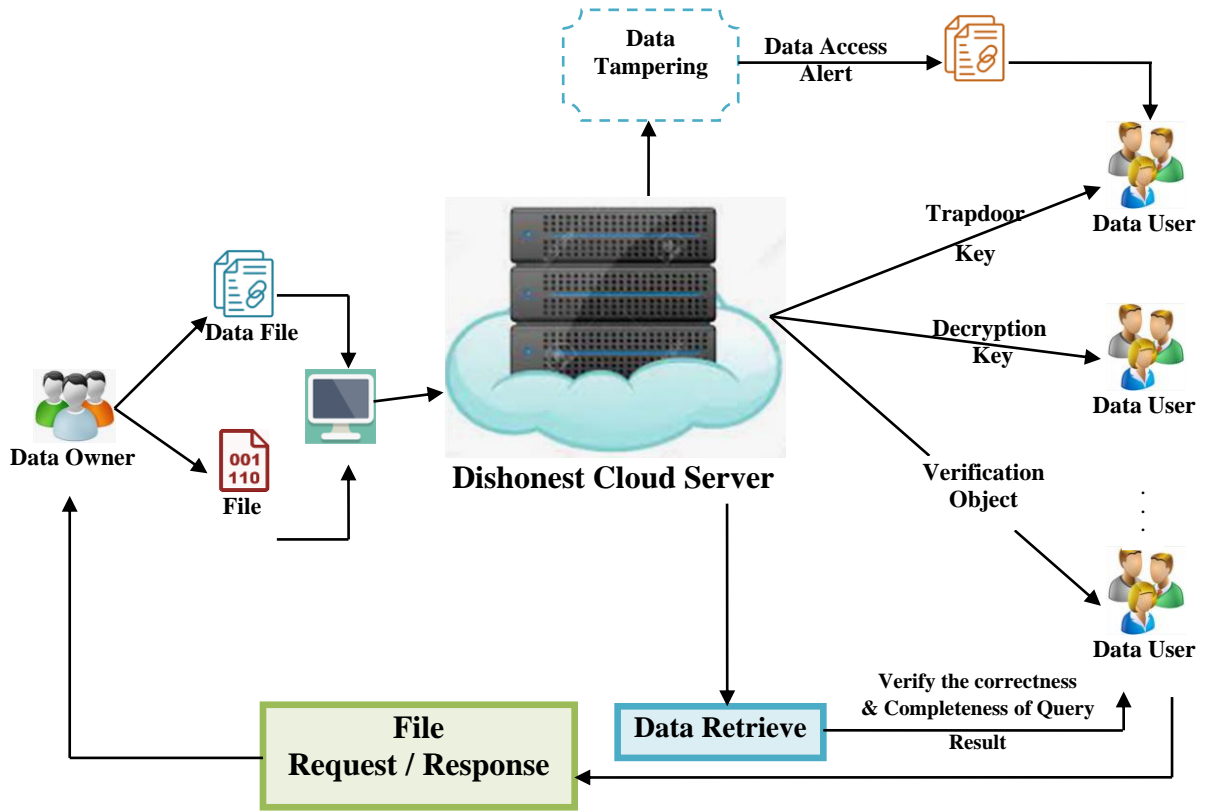

**Fig. 1 Technical Architecture of Outsourced Analysis of Encrypted Data**

To obtain the following attractive data security, GOOSE uses cryptosystem and secure multi-party computation: (i) no cloud node could indeed gain knowledge of the graph; (ii) no cloud node could indeed simultaneously learn the query and the query responses, and (iii) an outside network spectator could indeed gain knowledge of graph; the query; or the query answers. The core of the W3C's SPARQL 1.1 specification, Unions of Conjunctions of Regular Path Queries (UCRPQ), is supported by GOOSE as a query language and recursion queries. They demonstrate that the latency associated with cryptographic techniques scales linearly with input and output sizes.The FHE-based technique for mathematical systems will need to be demised at many levels. Re-encryption to preserve the usefulness of encrypted data [10,15]. Larger cipher messages and substantial processing expenses are needed for this. On the other side, the data owner wishes to control and analyze the growing customer data using public cloud services [50].

This study considers calculations for security savings for one essential chart examination: diagram extraterrestrial analysis for offshore charts in the cloud. The main task: Multiple information extraction procedures also depend on the Eigen decomposition of large frameworks. We consider a cloud-driven design with personal identification, content providers, and cloud suppliers as three synergistic groups. Charts are referred to as frameworks; their parts are stored and assembled by dispersed clients [17]. The information proprietor subsequently collaborates with cloud part initiatives to drive creepy investigation while safeguarding data security against the reputable but enquiring cloud service. While computations are made according to data contributors and proprietors.

## 3. Proposed Methodology

Using the SDHE and ASHE methodologies within the cloud-centric architecture presents several obstacles, which our research tackles. (1) Since SDHE permits homomorphic multiplication solely on a single level, implementing cloud-side operations is simple. However, the full extent of their costs has yet to be discovered. (2) ASHE techniques have smaller cipher text sizes, making storage and transmission effective. However, in the cloud, data providers must acquire, decode, and analyze information locally to ensure computational anonymity, as shown in Fig. 1.

We determine the privacy risk associated with sending sparse graph matrices and create a productive local differential—a secret technique for adding fictitious edges with identically encrypted values. Both may be rebuilt and adapted to a cloud infrastructure to accomplish practical-based division. The real effort of separating the consumer and cloud parts protects the confidentiality of data and analytic output, as shown in Fig. 2.

Protected search enables authorized data users to seek through the encoded data of the data owner and privately offers anonymized search terms [6,18,20,21]. It is a compelling adaptation of conventional cryptography for the cloud computing environment and is fueled by efficient content recovery from encrypted cloud data that has been subcontracted [16, 22-26]. A significant amount of study has been done on safe search terms and difficulties in cloud technology, with the goals of consistently enhancing search effectiveness, lowering computation and communication costs, and enhancing the range of search features with greater privacy and security protections [51]. All of these. strategies share the fundamental presumption that perhaps the cloud is an "honest-but-curious" phenomenon thatconsistently maintains resilient and reliable software and hardware environments.

As a consequence, whenever a search is finished, the cloud provider consistently returns accurate and comprehensive search queries without exception. For safe keyword search over secure cloud data, we officially present the provable secure searching model of the system and threat model and construct a perfectly all-right search outcomes classification method. We suggest a quick signature method based on public-key cryptography without certificates to validate objects' veracity.

### *3.1 Added Substance Homomorphic Encryption*

The following characteristic of added ingredient homomorphic encryption exists. The additive homomorphic procedure is shown as follows for two integers.

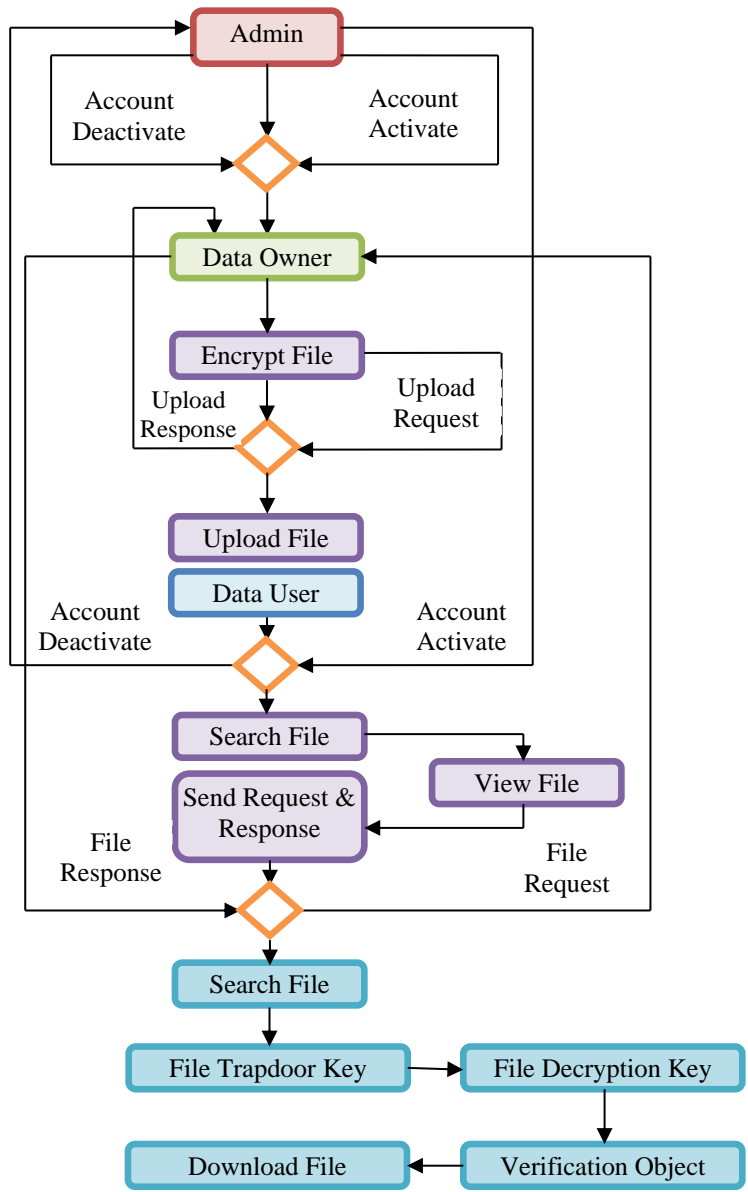

**Fig. 2 Process Flow of Proposed Approach**

$$E_n(x + y) = E_n(x) + E_n(y) \qquad (1)$$

We will utilize Paillier cryptography as an example of one of the most effective ASHE strategies to illustrate our ASHE-based procedures. A series of pseudo-homomorphic processes that form the basis of our procedures are made possible by additive homomorphic encryption. Unencrypted for one parameter, or either, we obtain

$$E_n(xy) = \sum_{i=1}^{y} E_n(x) = \sum_{i=1}^{x} E_n(y) \qquad (2)$$

$E_n(xy) = E_n(x)^y \mod P_k^2$, where $P_k$ is the public key, provides a more effective method to multiply for Paillier cryptography. Since an operand isn't encoded, we refer to it as pseudo-homomorphic multiplication. We can deduce the pseudo-homomorphic dot product, matrix-matrix multiplication (MMM), and matrix-vector multiplication (MVM), each employing one unencrypted parameter, from all these two essential aspects. Protecting the unencrypted operand is the main difficulty faced by ASHE-based solutions.

### 3.2 Some Degree Homomorphic Encryption

In recent years, systems for some degree of homomorphic encryption (SDHE) have indeed been created to accomplish a level or more of homomorphic multiplier concurrent with ASHE. For instance, it is possible to determine on encrypted numbers $E(n_i)$ while decoding them the sums $(n_1 + n_2)(n_3 + n_4) + (n_5 + n_6)(n_7 + n_8)$. Keep in mind that each value only requires one multiplication. In comparison, the multiplication in $n_1, n_2, n_3$ occurs twice. The degree-2 functions are frequently computed homomorphically using the SHE methods. Several well-known SHE prelisting: the BGN strategy, utilizing group pairings with elliptic curves, the RLWE method, relying on the ring learning-with-error issue; and the Catalano et al. [11] strategy, focused on an adaptation of the AHE strategy.

We will employ the RLWE method in the analysis instead of the other two because of cost concerns [28-33]. The decryption of the BGN technique relies on processing a discrete log, which has an $O(\sqrt{q})$ cost for unencrypted variables in the [0, q] range. We discover that it takes more than one second to decode 20-bit data by using the component dlog brute force technique, which may undoubtedly be reduced with some adjustment. The ciphertext extension of the Catalano et al. [11] algorithm led to its exclusion. When an N-dimensional space and an N*N-encoded matrix are multiplied, the result will contain $O(N^2)$ encoded components, which are too pricey to be sent to the customer. We omit the specifics among these techniques owing to space constraints.

**Algorithm - Privacy-preserving - (PP) sparse submission ($H_s$, $D_p$, $An_{a,b}$).**
**Input:**
$H_s$ - histogram,
$D_p$ - parameter (differential privacy),
$An_{a,b}$ - precise node degree.

Determine the bin containing $An_{a,b}$, where $Up_a$ and $Lo_a$ are its upper and lower bounds.

$x \leftarrow (Up_a - Lo_a)/D_p$;
$y \leftarrow x * 3.9$; // for y ≈ 3.9 for x = 1 the y scales linearly with x: y ≈ 3.9x;

Generate a variable $\phi_{a,b}$ based on dispersal Laplace (0, x); $K_{a,b} \leftarrow |y| + \phi_{a,b}$; add $An_{a,b}$ inadequate encryption and actual references to the listing; arbitrarily choose $K_{a,b}$ edges away from the others $N - An_{a,b}$ edges and as the encoded zero bits, encrypt it; Therefore, provide index (a, b) of the items for b ≥ a if the graph is directionless; if not, submit all $An_{a,b} + K_{a,b}$ items.

To create a brand-new Parlier Encryption-based verification object request method in which the Cloud provider has no idea whatever information the user is seeking or even which certification items will be presented to the user. To assess the precision and effectiveness of our suggested system, we offer comprehensive security specification and verification, as well as carry out thorough performance trials.

### 3.3 Query Process

The data user can validate the findings using the query result verification mechanism. In this article, we created a secure, straightforward to combine by providing a specific query result set. If the accumulation somehow fails to return either the number of or whichever qualifying files, the search client can do further checks and verify the accuracy of each data source in the accumulation [34-38]. This is known as a fine-grained query results validation mechanism.

The cloud computing idea enables speedy deployment and distribution of a shared pool of reconfigurable computational power, such as networking, processors, memory, programs, and applications, with minimum administrative labor or service provider participation.

Three separate keys will instantly be produced for the encrypted format when the data owner transmits to a remote server. To secure the anonymity of the validation objects while minimizing space and communication costs. Keys for trapdoors, verification objects, and decryption are generated automatically. The trapdoor key distinguishes between data owners and hackers [5,27,39-42,46]. The query results group and related validation objects are returned after a query is complete and are provided to the querying user, who uses the validation item to check the accuracy and comprehensiveness of the query results. Our suggested query outcomes validation approach allows the information to execute completeness verification before decoding search queries rapidly and verifies each encrypted data file in the query results set by the query client.

When a cloud server or other unauthorized party accesses information or data that the user has stored. Anytime someone tries to access the information or data, the data user will receive a warning. We may stop unauthorized users from obtaining user data or information by validating the verification object [43-45,49]. When the data held within them cannot be retrieved typically, data recovery is the act of saving (retrieving) unavailable, stolen, distorted, corrupted, or reformatted information from secondary stores, portable media, or files. We can still retrieve the entire document even if a hacker has access to the data or tampers with it.

## 4. Experimental Evaluations

We have demonstrated that, given the framework's supposition, all strategies that have been constructed ensure privacy. The tests will assess different expenses related to these strategies to determine which algorithms are more effective. Our analysis has three main components: evaluating the Search complexity in carrying out the ASHE and SDHE-based privacy-preserving (ii) the Query time for the cloud and data providers with various cryptographic techniques.

### 4.1. Setup

After the data owners' logins have been verified and granted access, the datasets will be uploaded to the cloud so everyone can access them. After choosing the file from the system, we must enter the date for the cloud system upload. We have a search function that may show the encrypted search, the uploaded time, the owner of the data, and the action. So that we can find out who submitted the file, its owner, and when it was posted, since there is action here, we must request the individual who submitted the file. When you request files from a user, they will respond with whether you can retrieve those files or not. The specific individual cannot access the item from the repository if the owner does not grant access. If the user grants access to the file, the person can take any action they require. The four essential components are file name, user name, timestamp, state, and activity. The state will be given, and the activity will be a document action allowed just after the owner has provided the authorization.

### 4.2. Storage Complexity

When compared to existing methods like Dynamic Searchable Symmetric Encryption [9], Linearly-Homomorphic Encryption [11], and SPARQL [14], our schemes' storage complexity is $O(N^2)$ and $O(N + \lambda)$, correspondingly.

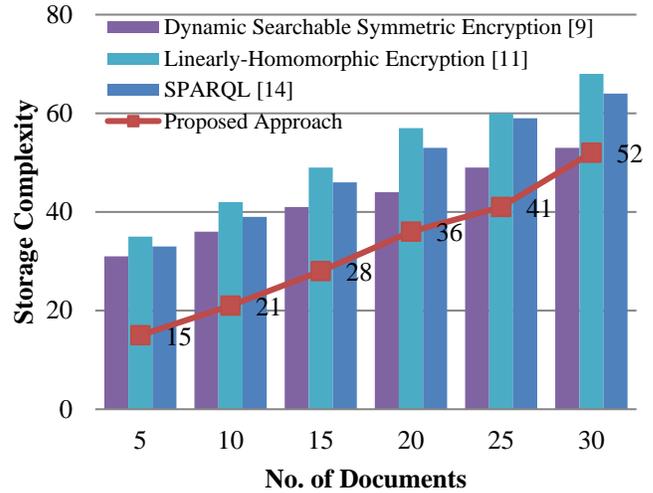

**Fig. 3 Comparison of Storage complexity**

In reality, the needed level of protection is supposed to be indeed achieved with a high enough security parameter λ level. Even though there are 230 documents, as shown in Fig. 3, the complexity of storage is reduced when we select $\lambda = 2^{64}$ and $\lambda = 2^{80}$ in the proposed strategies. The SPARQL schemes keep the data on the searching consumer and the cloud server, which could result in expensive storage costs for search users. While the majority of the data in Linearly-Homomorphic Encryption and Dynamic Searchable Symmetric Encryption is sent to a cloud server, which can be integrated into massive storage, the cost of the searching customer is little since only the numbers gathering D is kept there.

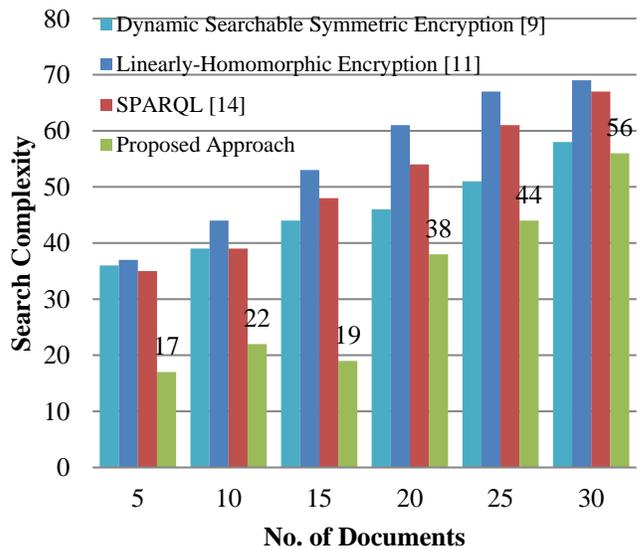

**Fig. 4 Comparison of Search complexity**

*4.3. Search Complexity*

Moreover, the search overhead of the proposed strategies, as well as the SPARQL strategies [14], is $O(N^2)$ and $O(\lambda \cdot \log^2 N)$, accordingly. Even if there are $2^{30}$ documents, as depicted in Fig. 4, our strategies are less search-complex than SPARQL strategies.

*4.4. Query Time*

The length of the dictionary and the number of documents significantly impact the calculation cost in the query phase, as illustrated in Fig. 5.

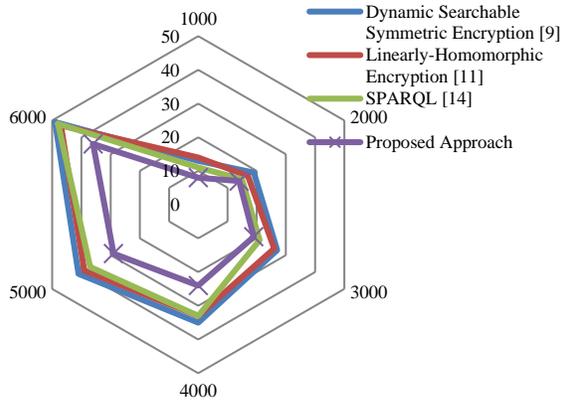

**Fig. 5 Comparison of Query time**

In contrast, the amount of query terms essentially has no effect. The strategies could be effective during the query stage as well. Our methods reduce storage complexity, modifying sophistication, and difficulty associated with creating indexing, a trapdoor, and a search. Exceptionally, compared to other systems, the upgrading complexity of our approaches may be nearly nonexistent.

## 5. Conclusion

We develop a platform for the spectrum analysis of huge matrices while maintaining privacy, which offers solid confidentiality assurance defense against sincere but inquisitive cloud providers. Secured graph data can be uploaded to the cloud by data contributors, and Using secure protocols, the analysis is conducted between the data owner and the cloud. The system successfully restricted in-house analyses to the resource-restricted data owner and storage capacity and safely outsourced the expensive analyses to the cloud. We create two privacy-preserving strategies for spectrum analyzers and investigate how they are built using additive substance homomorphic encryption (ASHE) and some degree of homomorphic encryption (SDHE).

The plaintext operands of the AHE methods must be protected from attackers, so we created masking approaches that fulfill the needed privacy guarantees while enabling the data owner to increase complexity. Large sparse matrices aid the privacy-preserving approach. We created the privacy-preserving dense data submission methodology for resource providers to find a balance between sparse data and anonymity. The approach to data sparsity dramatically lowers costs for the data owner. Using ciphertext packing in the RLWE-based approaches reduces computation overhead, while the Paillier-based methods significantly reduce online storage and data proprietors' transmission losses.

In the future, the cloud will need to seek across the complete database. It is highly wasteful and renders the technique of outsourced data-Search worthless. Future research in this field will focus on improvements for the effective verification of vast amounts of data that have already been outsourced. This technology currently only operates in partially authorized clouds, but it will eventually be expanded to include all cloud settings and can offer higher security. Additionally, we can expand our search approach in the future to employ external devices while protecting confidentiality.